\documentclass[11pt, a4paper]{article}

\usepackage{jheppub}
\usepackage{amsmath,amssymb}
\usepackage{epsfig,graphicx}
\usepackage{subfigure}
\usepackage{graphicx}
\usepackage{rotating}
\usepackage{cancel}
\usepackage{bm}
\usepackage{color}
\usepackage{comment}
\usepackage{psfrag}

\newcommand{\exclude}[1]{}

\def\beq{\begin{equation}}
\def\eeq{\end{equation}}

\preprint{IPPP/15/39, DCPT/15/78}

\begin{document}
\title{Probing MeV to 90 GeV axion-like particles with LEP and LHC}

\author{Joerg Jaeckel$^1$ and Michael Spannowsky$^2$}
\affiliation[1]{Institut f\"ur theoretische Physik, Universit\"at Heidelberg, Philosophenweg 16, 69120 Heidelberg, Germany}
\affiliation[2]{Institute for Particle Physics Phenomenology, 
  Durham University, Durham DH1 3LE, United Kingdom.}

\abstract{Axion-like particles (ALPs), relatively light (pseudo-)scalars coupled to two gauge bosons, are a common feature of many extensions of the Standard Model. 
Up to now there has been a gap in the sensitivity to such particles in the MeV to 10 GeV range. In this note
we show that LEP data on $Z\to\gamma\gamma$ decays provides significant constraints in this range (and indeed up to the $Z$-mass). 
We also discuss the sensitivities of LHC and future colliders. Particularly the LHC shows promising sensitivity in searching for a pseudo-scalar with $4 \lesssim m_a \lesssim 60$ GeV in the channel $pp \to 3 \gamma$ with $m_{3\gamma}\approx m_{Z}$.}

\maketitle

\section{Introduction}
Over the last few years there has been a rising interest in searching for particles with low mass but also weak coupling to the Standard Model.
In part this is motivated by the simple fact that ``the new particles have weak couplings to the Standard Model'' provides an equally good answer to the question ``why haven't we found the new physics'' as ``the new particles are very heavy''. Accordingly we should search in both of these directions.
Additional motivation comes from theoretical studies demonstrating that relatively light and weakly coupled particles arise quite naturally in a wide range of extensions of the Standard Model, also in connection with dark matter. See, e.g.~\cite{Jaeckel:2010ni,Essig:2013lka,Alekhin:2015byh} for some reviews/overviews. 

In this note we will be concerned with lightish (pseudo-)scalar particles, called axion-like particles or ALPs~\cite{Masso:1995tw,Masso:1997ru,Masso:2004cv,Jaeckel:2006xm}.
ALPs are loosely defined as relatively light scalar or pseudoscalar particles coupled to two gauge bosons and/or two Standard Model
fermions\footnote{Their name, of course, originates from the similarity to the famous axion introduced to solve the strong CP problem~\cite{Peccei:1977hh, Peccei:1977ur, Wilczek:1977pj, Weinberg:1977ma}. The axion crucially features a low mass and couplings to two gauge bosons as well as optional couplings to the Standard Model fermions. As a consequence of solving the strong CP problem, the axion couplings are proportional to the axion mass with an essentially known and fixed proportionality constant. In any plot of mass vs coupling strength axion models therefore populate a relatively narrow band. From the phenomenological point of view we can simply take axion-like particles as a straightforward generalization of axions where we relax the strict relation between mass and coupling and allow them to populate the whole mass vs coupling plane.}. 
They are one of the prime test-models in the search for light weakly coupled new physics. 
In field theory models (pseudo-)scalars with such interactions naturally arise as pseudo-Nambu-Goldstone bosons of spontaneously broken approximate symmetries~\cite{Masso:1995tw} and or by mixing with the Higgs boson~\cite{Branco:2011iw,Broggio:2014mna,Dolan:2014ska}. In string models moduli and string axions provide natural candidates~\cite{Witten:1984dg,Conlon:2006tq,Svrcek:2006yi,Arvanitaki:2009fg,Acharya:2010zx,Cicoli:2012sz}.
Last but not least, ALPs could also provide nice messengers to dark matter (sectors)~\cite{Freytsis:2010ne,Dienes:2013xya} and in some cases could even be the dark matter particles themselves~\cite{Masso:2004cv,Arias:2012az}.

In this note we want to focus in particular on axion-like particles whose dominant interaction with the Standard Model is via two gauge bosons (i.e. where interactions with the Standard Model fermions can be neglected). This can be viewed as a simple test example\footnote{Therefore we also happily continue the coupling to relatively large values that may be difficult to generate in a perturbative embedding.}, but such a situation also arises quite naturally
in string models~\cite{Cicoli:2012sz}. Practically we consider interactions with two photons and with two hypercharge bosons,
\begin{equation}
\label{bothinteractions}
{\mathcal{L}}_{\rm int}\supset -\frac{1}{4}g_{a\gamma\gamma}aF^{\mu\nu}\tilde{F}_{\mu\nu}\quad{\rm or}\quad -\frac{1}{4}g_{aBB}aB^{\mu\nu}\tilde{B}_{\mu\nu}.
\end{equation}
This interaction is for the specific case of a pseudo-scalar, but the analyses of this paper can be straightforwardly generalized to
scalar, and we expect quantitatively similar results.

\begin{figure}[!t]
\centering
\includegraphics[width=0.45\textwidth]{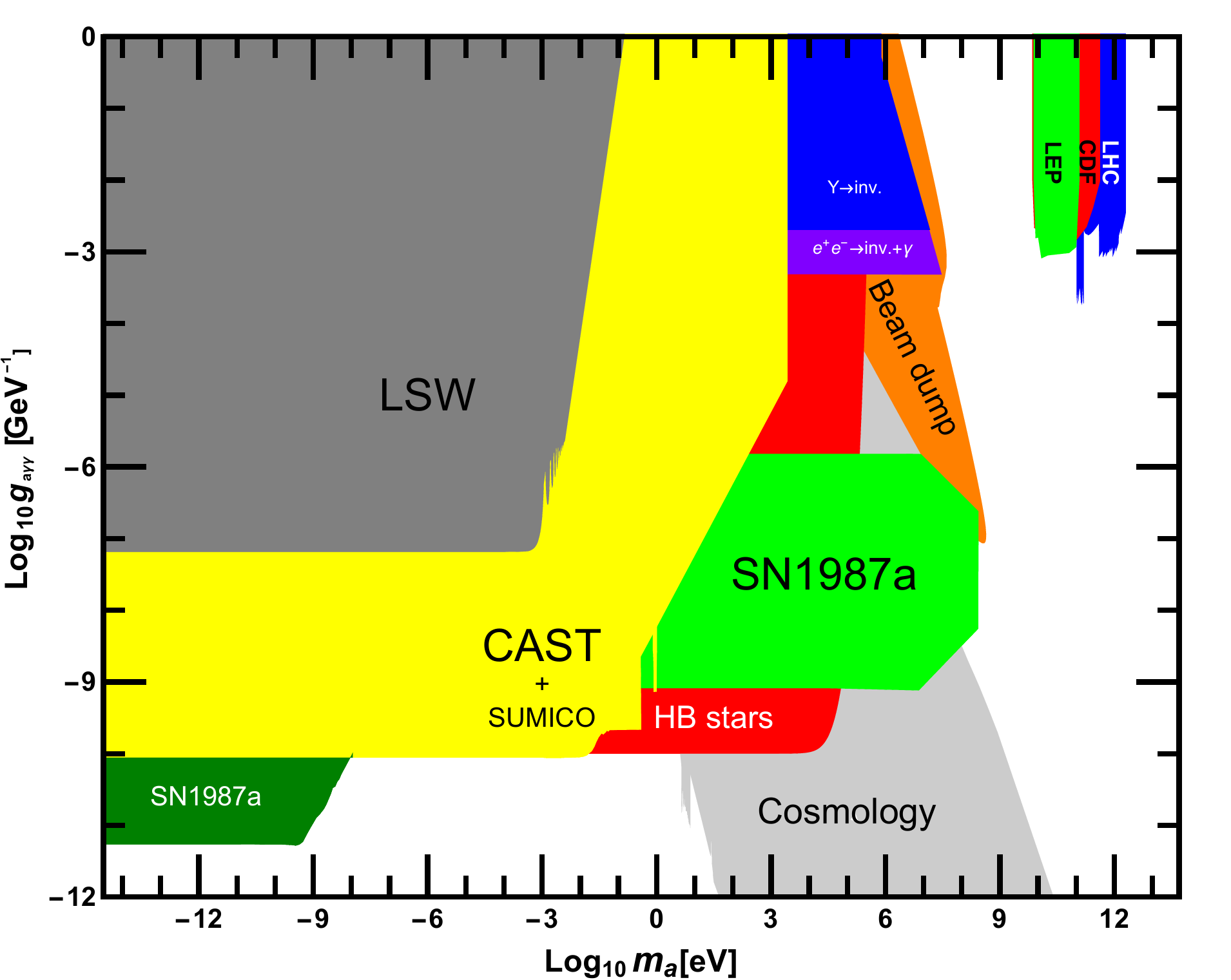}
\hspace*{1cm}
\caption{Limits on the axion-like particle to two photon coupling. Figure slightly adapted from~\cite{Alekhin:2015byh} which is a compilation adapted from~\cite{Redondo:2008en,Jaeckel:2010ni} updated with~\cite{Cadamuro:2011fd,Hewett:2012ns,Jaeckel:2012yz,Mimasu:2014nea,Payez:2014xsa,Millea:2015qra}. Note the gap in the MeV to 10 GeV region.}
\label{current}	
\end{figure}

This note is motivated by two very simple observations:
\begin{itemize}
\item{} The striking gap in the limits on the ALP coupling to two photons shown in Fig.~\ref{current} in the MeV to roughly 10 GeV 
region\footnote{Although future fixed target experiments such as, e.g. SHiP~\cite{Anelli:2015pba}, will nicely extend~\cite{Alekhin:2015byh} the ``beam dump'' region in Fig.~\ref{current}, they are limited to weaker couplings
because they require a sizeable decay length for the ALPs.}.
While reactor experiments have explored this mass region~\cite{Vuilleumier:1981dq,Zehnder:1982bg,Datar:1982ef,Alekseev:1982iq,Cavaignac:1982ek,Ananev:1983ki,Ketov:1986az,Koch:1986aq,Chang:2006ug} the corresponding limits depend on a coupling to fermions (typically nucleons) and are not directly applicable if there only is a coupling to gauge bosons.
\item{} The interaction with hypercharge bosons can to some degree be viewed as the more fundamental one. However, this coupling allows a decay
of a $Z$-boson via $Z\to a\gamma$. It is therefore natural to look for unusual decays of the $Z$ which promises sensitivity to ALPs with mass $\lesssim m_{Z}\approx 90~{\rm GeV}$.
\end{itemize}

\begin{figure}[t]
    \begin{center}
    \subfigure[]{
\begin{picture}(168,110)(-20,0)
\includegraphics[width=6cm]{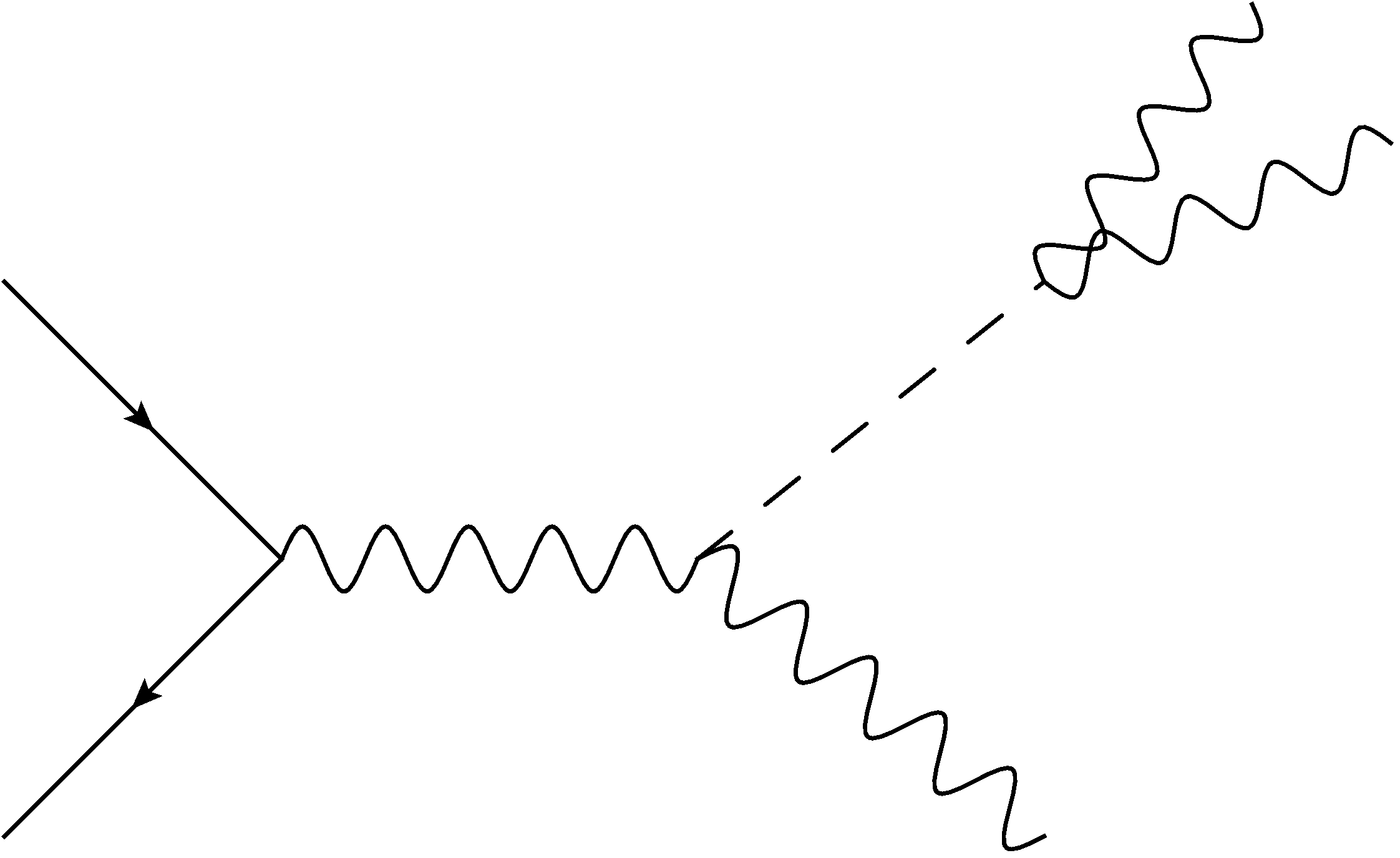}
\put(-115,20){$\displaystyle \gamma^{\star}$}
\put(-74,10){$\displaystyle \gamma$}
\put(-74,55){$\displaystyle a$}
\put(-20,70){$\displaystyle \gamma$}
\put(-34,100){$\displaystyle \gamma$}
\put(-155,60){$\displaystyle e^{-}$}
\put(-155,8){$\displaystyle e^{-}$}
         \end{picture}
\label{virtualgagaga}}
         \hspace*{1cm}
                  \subfigure[]{
                  \begin{picture}(168,110)(-20,0)
\includegraphics[width=6cm]{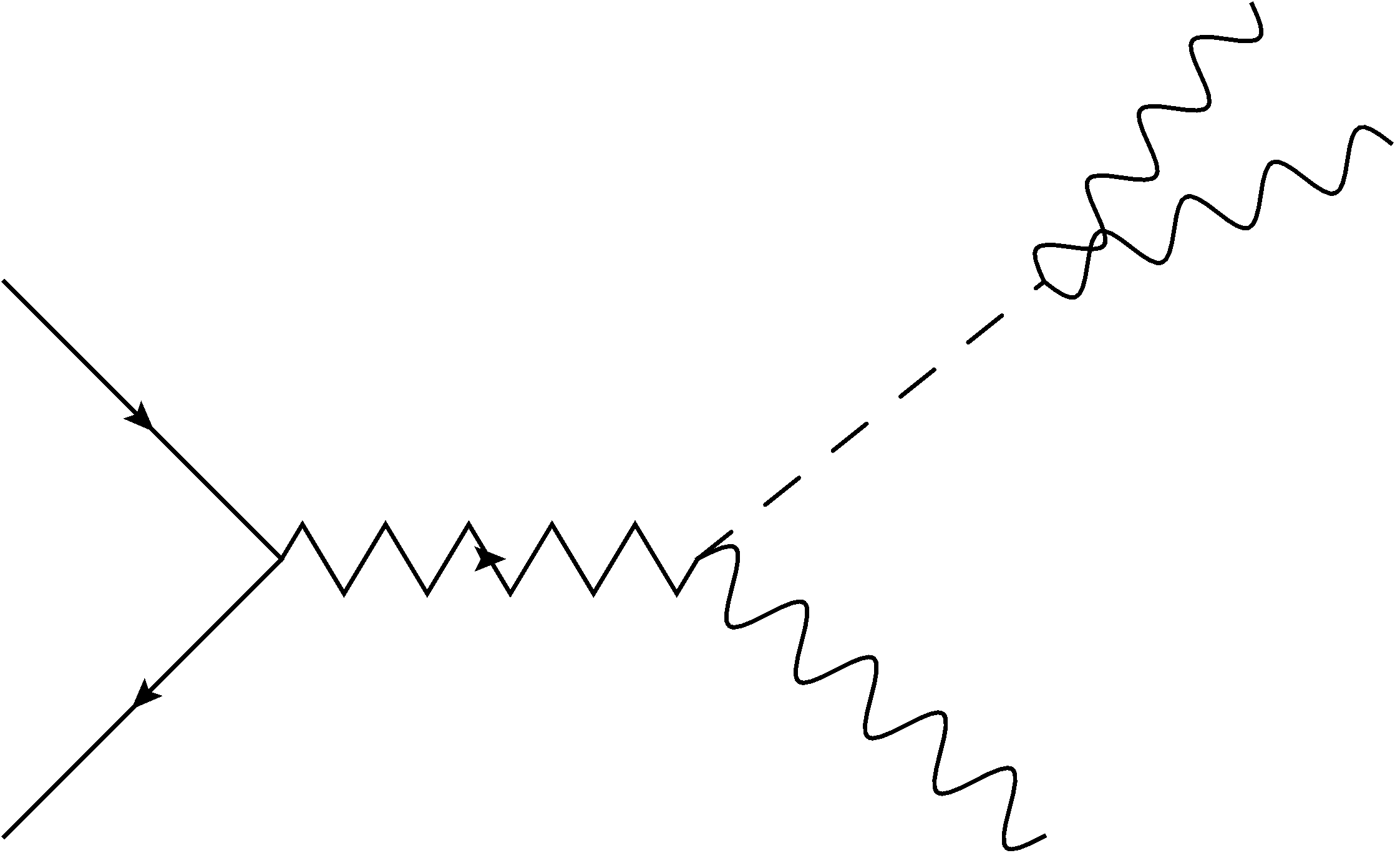}
\put(-135,20){$\displaystyle Z^{\rm (on-shell)}$}
\put(-74,10){$\displaystyle \gamma$}
\put(-74,55){$\displaystyle a$}
\put(-20,70){$\displaystyle \gamma$}
\put(-34,100){$\displaystyle \gamma$}
\put(-155,60){$\displaystyle e^{-}$}
\put(-155,8){$\displaystyle e^{-}$}
         \end{picture}
\label{Zgagaga}}
        \caption{Production of ALPs with subsequent decay into two photons. {\bf Left panel:} $a+\gamma$ production via a virtual photon and subsequent decay to $3\gamma$. {\bf Right panel:} production of an on-shell $Z$ and subsequent decays into $a+\gamma$ and then $3\gamma$. }
\label{feyman}
    \end{center}
\end{figure}
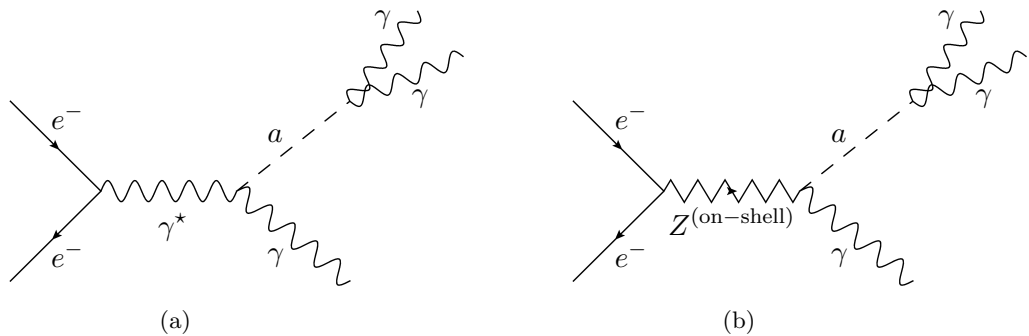

Indeed LEP data~\cite{Acciarri:1994gb,Anashkin:1999da} on the decay $Z\to3\gamma$ has already been used to constrain ALP couplings to two photons~\cite{Mimasu:2014nea} via the
process $e^{+}e^{-}\to a+\gamma, \,\,a\to 2\gamma$ as shown in Fig.~\ref{virtualgagaga}.
Our analysis differs in two essential points. First we also consider the coupling to the hypercharge bosons. This allows for the decay of an on-shell $Z$ into an ALP and a photon in contrast to the production of an ALP and a photon via a (highly) virtual photon (cf. Fig.~\ref{Zgagaga}). This on-shell production significantly enhances the sensitivity. Second and more importantly we also use data on $Z\to 2\gamma$ decays. The search for a $3\gamma$ signature is only sensitive for sufficiently high ALP masses such that the two photons arising from the decay of a fast moving ALP can be separated.
In practice this limits the sensitivity to masses $m_{a}\gtrsim 10\,{\rm GeV}$. However two tightly collimated photons essentially produce the same detector response as one photon with the combined energy. One can therefore use the $Z\to 2\gamma$ search in this regime.
This allows us to fill in the sensitivity gap in the MeV to 10 GeV region as shown in our result plots Figs.~\ref{resultleft} and \ref{resultright}.
Finally, we note that we only use data from the Z-pole measurement~\cite{Akrawy:1990zz,Abreu:1991rm,Abreu:1994du,Acciarri:1994gb,Acciarri:1995gy,Agashe:2014kda}, while~\cite{Mimasu:2014nea} also uses higher energy data~\cite{Anashkin:1999da}. 

The remainder of this note is structured as follows. In the next Section~\ref{lepsearch} we describe our analysis of the two and three photon LEP searches and present the corresponding new limits. Following this we have a look at the prospects at LHC and at future colliders in Section~\ref{future}.
We briefly summarize and conclude in Section~\ref{conclusions}.

\section{Searching ALPs at LEP with $Z\to2\gamma$ and $Z\to 3\gamma$}\label{lepsearch}
\subsection{ALPs from $Z$ decays}
For some time of its operation LEP has run with an energy on or close to the $Z$ mass. During this
time a huge number of $Z$ were produced ($\sim {\rm few}\times 10^6$). This opportunity has been used
to constrain the branching ratios for unusual $Z$ decays, in particular decays to two and three photons~\cite{Akrawy:1990zz,Abreu:1991rm,Abreu:1994du,Acciarri:1994gb,Acciarri:1995gy,Agashe:2014kda}.

In presence of a coupling of ALPs to two hypercharge bosons, the $Z$ boson can decay to an ALP and a photon with a rate,
\begin{equation}
\Gamma_{Z\to a+\gamma}=\frac{g^2_{aBB}\sin^2(\theta_{W})\cos^2(\theta_{W})}{96\pi}m^{3}_{Z},
\end{equation}
where $\theta_{W}$ is the Weinberg angle.\\

The ALP subsequently decays into two photons with a rate,
\begin{equation}
\Gamma_{a\to2\gamma}=\frac{g^{2}_{aBB}\cos^{4}(\theta_{W})}{64\pi}m^{3}_{a}.
\end{equation}

For detection purposes two factors are important. First the decay length in the laboratory frame has to be within the detector, more precisely before the
electromagnetic calorimeter. It is given by
\begin{equation}
\ell_{\rm decay}=\frac{\gamma_{a}}{\Gamma_{a\to 2\gamma}}\approx \frac{m_{Z}}{2m_{a}}\frac{1}{\Gamma_{a\to 2\gamma}}\sim
2\,{\rm cm}\left(\frac{10^{-3} {\rm GeV}^{-1}}{g_{aBB}}\right)^2\left(\frac{100\,{\rm MeV}}{m_{a}}\right)^4,
\end{equation}
where $\gamma_{a}$ is the relativistic factor for the ALP.
On the left hand side we have used that in the LEP $Z$ peak measurement $Z$ bosons are produced at rest and consequently
for ALP masses much smaller than the $Z$ mass, half the energy goes into the ALP.
For small masses and couplings the decay length limits the sensitivity of the measurement. 
In the low mass regime we therefore only take the fraction of events that decays within the first 10~cm into account.

\begin{figure}[!t]
\centering
\includegraphics[width=0.45\textwidth]{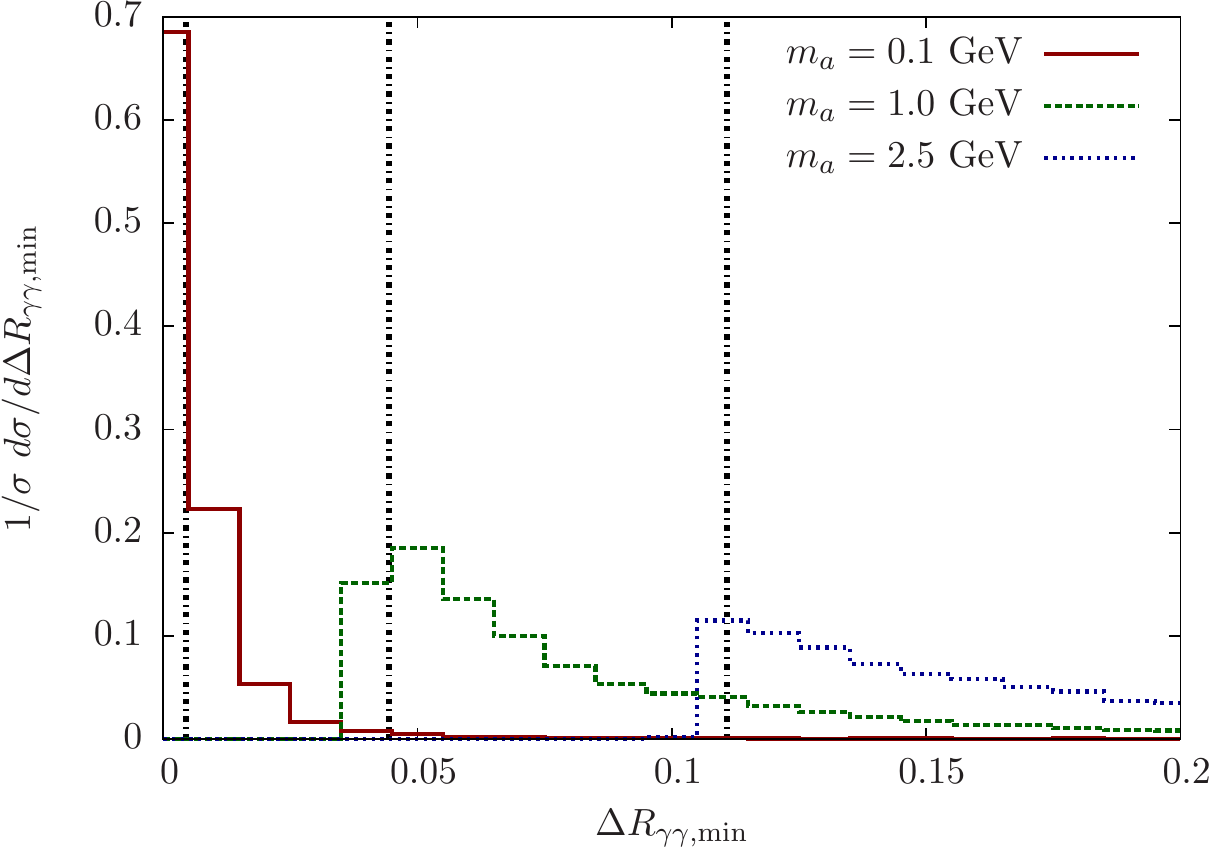}
\caption{$\Delta R$ separation of the two closest photons for different values of $m_a$ in the process $e^+e^- \to Z \to a \gamma \to 3 \gamma$. The black vertical lines correspond to $\Delta R = 4m_a/m_Z$. 
}
\label{fig:dR}	
\end{figure}

The second important factor is whether the two photons  of the $a\to 2\gamma$ decay can be separated in the experiment.
This decides whether the constraints of the $Z\to 3\gamma$ measurements are applicable or one has to consider those from $Z\to 2\gamma$.
For central production, the separation is roughly given by
\begin{equation}
\label{separation}
\Delta R\sim \frac{2 m_a}{p_T}\sim \frac{4 m_{a}}{m_{Z}},
\end{equation}
see Fig.~\ref{fig:dR}.

For example requiring a separation of 20 degrees as in~\cite{Acciarri:1994gb} $\Delta R\sim \sin(20^{\circ})\sim 0.34$ one finds that this limits the mass reach of the $Z\to 3\gamma$ to ALP masses $m_{a}\gtrsim {\rm few} \,{\rm GeV}$ (cf. also~\cite{Mimasu:2014nea}).

However, if the separation is very small the two photons from the ALP decay appear essentially as one photon of the combined energy.
Indeed already the LEP collaborations themselves used the $Z\to 2\gamma$ measurements to constrain the branching ratios
of $Z$ into photon and mesons which subsequently decay into two photons, e.g. $Z\to \gamma+\pi^{0}$ or $Z\to \gamma+\eta$.

Let us now apply this to the case of our ALPs.
In practice we consider three regions.
\begin{itemize}
\item[1)]{} $m_{a}\leq m_{\pi^{0}}=135\,{\rm MeV}$. Here we use the limit from on the branching ratio $BR(Z\to\gamma+\pi^{0})\leq 5.2\times10^{-5}$~\cite{Acciarri:1995gy} simply as $BR(Z\to\gamma+a)=\Gamma_{Z\to a\gamma}/\Gamma_{Z}\leq 5.2\times10^{-5}$.
\item[2)]{} $m_{\pi^{0}}\leq m_{a}\leq 10\,{\rm GeV}$. In this region we have simulated angular distribution for the production and decay $e^{+}+e^{-}\to Z\to a+\gamma\to 3\gamma$. This we compared bin by bin to the distribution given in~\cite{Acciarri:1995gy}. For those bins where there was a (non-significant) excess in a bin we have added this excess to the statistical uncertainty to obtain a conservative bound.
\item[3)]{} $10\,{\rm GeV}\leq m_{a}\leq m_{Z}=91.2\,{\rm GeV}$ most of the decays result in clearly seprable $3\gamma$ events. Accordingly we have used the limit $BR(Z\to\gamma+a\to 3\gamma)\leq BR(Z\to3\gamma)\leq 10^{-5}$ from~\cite{Acciarri:1994gb}.
\end{itemize}

\begin{figure}[!t]
\centering
\subfigure[]{\includegraphics[width=0.48\textwidth]{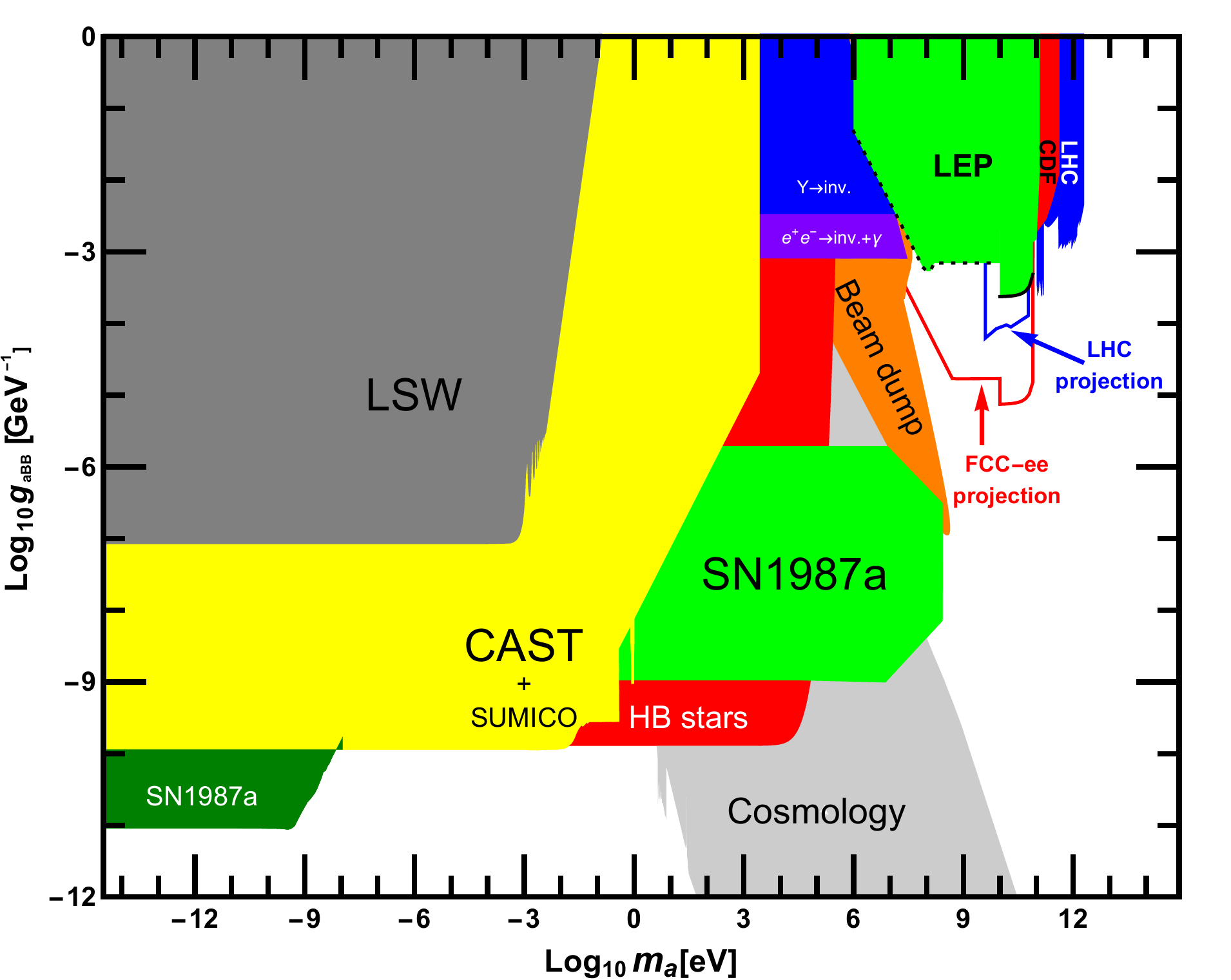}\label{resultleft}}
\hspace*{0.3cm}
\subfigure[]{\includegraphics[width=0.48\textwidth]{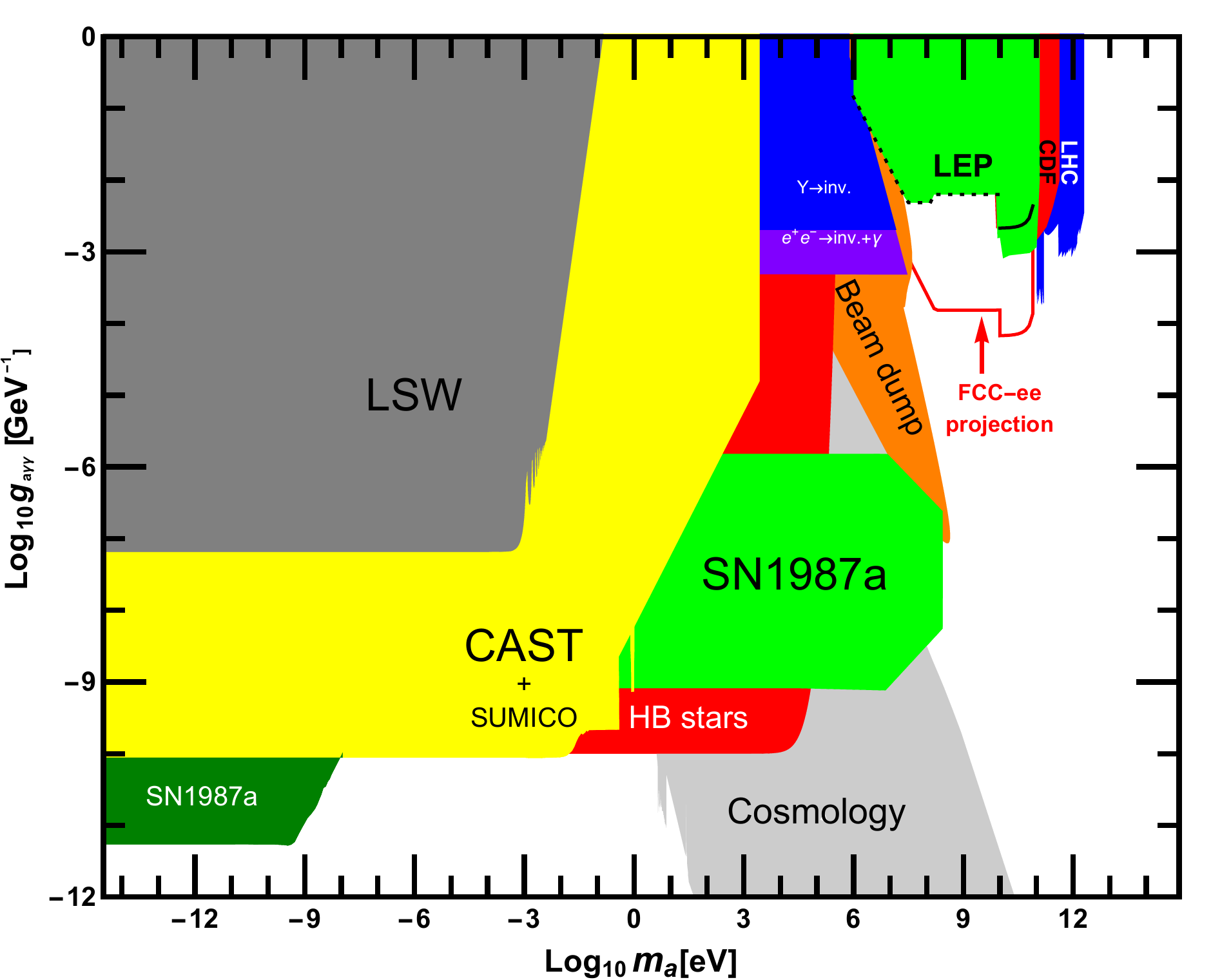}\label{resultright}}
\caption{{\bf Left panel:} Limits on a coupling to two hypercharge bosons. {\bf Right panel:} Limits on a coupling only to photons.
The new LEP limits from 2 and 3 photon signatures are shaded in green and enclosed by dashed and solid black lines, respectively. 
The future FCC-ee limit is indicated by the red solid line. Our projected LHC sensitivity for 13 TeV  and 100 fb$^{-1}$ by the blue line (only applicable to the coupling to hypercharge bosons).
The rest of the figure is
adapted from~\cite{Alekhin:2015byh,Redondo:2008en,Jaeckel:2010ni,Cadamuro:2011fd,Hewett:2012ns,Jaeckel:2012yz,Mimasu:2014nea,Payez:2014xsa,Millea:2015qra}.}
\label{result}	
\end{figure}

The resulting limits are shown in green in Fig.~\ref{resultleft}. The solid line indicates the limit from
$Z\to 3\gamma$ and the dashed one that from the $Z\to 2\gamma$ measurements. The latter one is sensitive to relatively low ALP masses
and indeed most of the ``hole'' mentioned in the introduction is covered by this measurement.

We note that the stronger limit at large masses from the $Z\to 3\gamma$ measurement as compared to Ref.~\cite{Mimasu:2014nea} arises from the on-shell production of the $Z$ boson. Using the production via a virtual photon, as done in the next subsection, produces results roughly compatible with Ref.~\cite{Mimasu:2014nea}.

\subsection{ALPs from production via virtual photons}
While a pure coupling to two hypercharge bosons will always lead to a decay $Z\to a+\gamma$ it is possible
to have a combination of couplings to two hypercharge bosons and to two ${\rm SU}(2)_{\rm weak}$ bosons such that the corresponding
ALP-photon-$Z$ coupling does not exist and one is effectively dominated by a two photon coupling. Although we think that the presence of an ALP-photon-$Z$ coupling is rather generic it is nevertheless worthwhile to also consider the latter case.

If ALPs are coupled only to two photons, production has to occur via a virtual photon as shown in Fig.~\ref{virtualgagaga}.
In general an ALP photon pair produced in this manner has an invariant mass given by the centre of mass energy of the two colliding particles producing the virtual photons and therefore not necessarily $m_{Z}$. However, this is still the case for
the LEP measurement at the $Z$ peak, because at a lepton collider such as LEP the two colliding particles have a definite energy given by the collider energy 
and for the measurements we consider this was (nearly) $m_{Z}$.

Aside from the difference in production the analysis follows along similar lines as in the previous subsection.
To obtain the limits in this case we have simply rescaled the limits with the appropriate lower ALP production cross section.

The resulting limits are shown in light green Fig.~\ref{resultright}. Again the solid line indicates the $Z\to 3\gamma$ measurement and the dashed one
the $Z\to 2\gamma$ limit. As above we see that the two photon measurement extends the reach to low masses. In the overlapping region our
limits are slightly weaker than those of~\cite{Mimasu:2014nea} which also used data based on more integrated luminosity at energies off the Z-peak (since the production via photons is always off-shell there is no special benefit in Z-peak data). 

\section{ALPs at LHC and Future Colliders}\label{future}
\subsection{Future electron-positron machines}
Let us first consider the sensitivity of future lepton colliders such as ILC~\cite{Behnke:2013xla,ILCweb}, CEPC~\cite{CEPCweb}, and FCC-ee~\cite{Gomez-Ceballos:2013zzn,FCCweb}. 
For these the analysis that one can perform is exactly as in the previous section and limits can be obtained for both the pure photon
and the hyperacharge coupling in Eq.~\eqref{bothinteractions}.

Indeed with at FCC-ee running at the Z-peak we can hope for about $10^7$ times as many $Z$-bosons as were produced with LEP-I running at the $Z$-peak. Naively, we can scale the improvement in the branching ratio as $\sqrt{N_{Z}}$. We therefore expect that the branching ratios could be  improved by a factor $10^{3}-10^{5}$. Accordingly the limits on the couplings are improved by a factor of $30-100$. This is shown by the red lines labelled FCC-ee indicated in Fig.~\ref{result}. Expecting improvements also in the detectors and a dedicated analysis one can hope that the actual sensitivity will actually be significantly better.

\subsection{Testing the hypercharge coupling with LHC}
However, while the FCC-ee or any similar machine is still in the distant future the LHC is running right now.
Searches for ALPs coupled to two gauge bosons at LHC have already been considered in~\cite{Jaeckel:2012yz}. 
However, in that paper the opportunity to look for relatively low mass ALPs from the decay of $Z$ bosons via a coupling of the ALP to hypercharge bosons was not considered.
This is what we will do here. As we will see this allows significant improvements of the LHC sensitivity for ALPs coupled to hypercharge 
bosons\footnote{For a study of relatively light pseudo-scalars coupled to fermions at LHC see~\cite{Kozaczuk:2015bea}.} (second coupling in Eq.~\eqref{bothinteractions}).

If one has the option of an on-shell $Z$-boson decaying into $a+\gamma$ one has an additional clean search channel: one can look for two or three photons reconstructing to the $Z$-mass. 
With a total cross section for $Z$ production in the range of
\begin{equation}
\sigma(pp\to Z+X)\sim 3\times 10^{4}\,{\rm pb}\quad  7\,\,{\rm TeV}
\end{equation}
LHC has already produced a very large number,
\begin{equation}
N^{\rm LHC}_{Z}\sim 7\times 10^8,
\end{equation}
of $Z$ bosons.

At LHC experiments, isolated photons are reconstructed by studying the shower profile in the electromagnetic calorimeter. Highly collimated photon-pairs are likely to violate the reconstruction requirements for isolated photons. For a first sensitivity study of this channel we will limit ourselves to the scenario where three isolated photons can be reconstructed. We parametrise the photon reconstruction efficiency and jet-fake-photon rate according to \cite{ATLASeff}, i.e. we smear the
momenta of all reconstructed final state objects with
Gaussians and parametrise the photon reconstruction efficiency with  
\begin{equation}
\mathcal{E}_\gamma = 0.76 - 1.98 e^{-p_{T,\gamma}/16.1~\mathrm{GeV}}
\end{equation}
and the jet-photon fake rate with
\begin{equation}
\mathcal{P}_{j\to \gamma}=0.0093 e^{-0.036p_{T,j}/\mathrm{GeV}}.
\end{equation}
Two of the photons in the signal tend to be highly collimated. Hence, we define photons to be isolated if $p_{T,\gamma} \geq 20$ GeV and if the amount of hadronic energy in a cone of $R = 0.1$ around the photon is less than $10\%$ of the photon's transverse energy. With three fairly hard photons the final state is likely to satisfy trigger requirements.

We generate signal samples with Madgraph \cite{Alwall:2011uj} and background samples with Sherpa \cite{Gleisberg:2008ta}. As dominant backgrounds we consider the processes $3\gamma$, $2\gamma+j$ and $\gamma + 2j$, where the first background is irreducible while the other two require one or two of the photons to be mis-identified.

\begin{figure}[!t]
\centering
\includegraphics[width=0.45\textwidth]{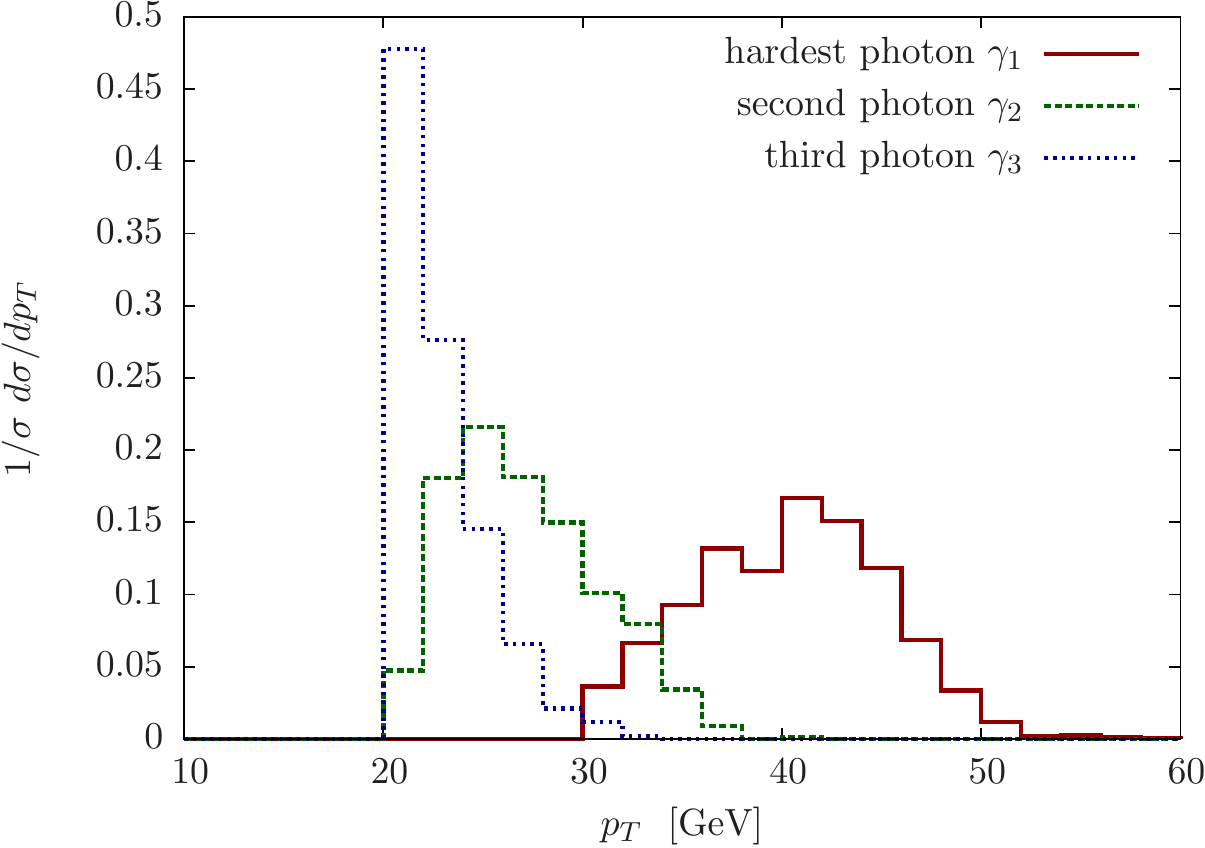}
\includegraphics[width=0.45\textwidth]{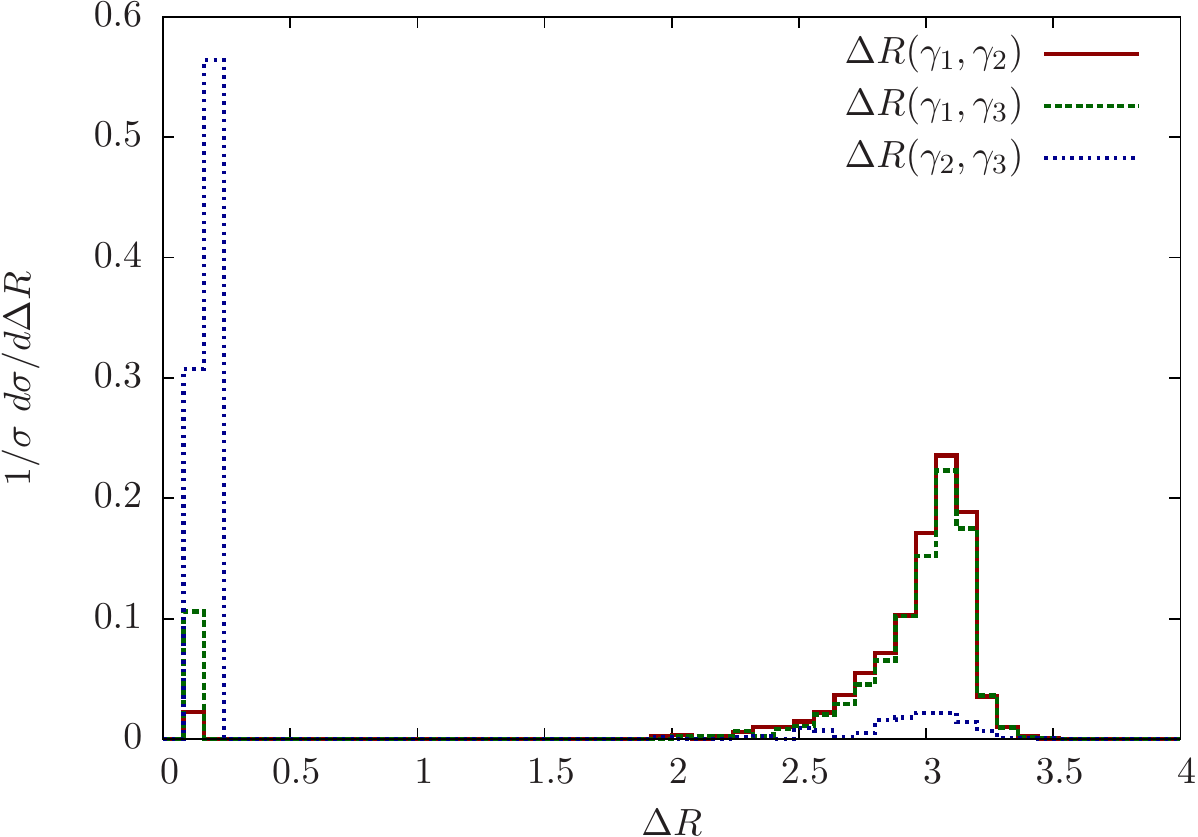} \\
\includegraphics[width=0.45\textwidth]{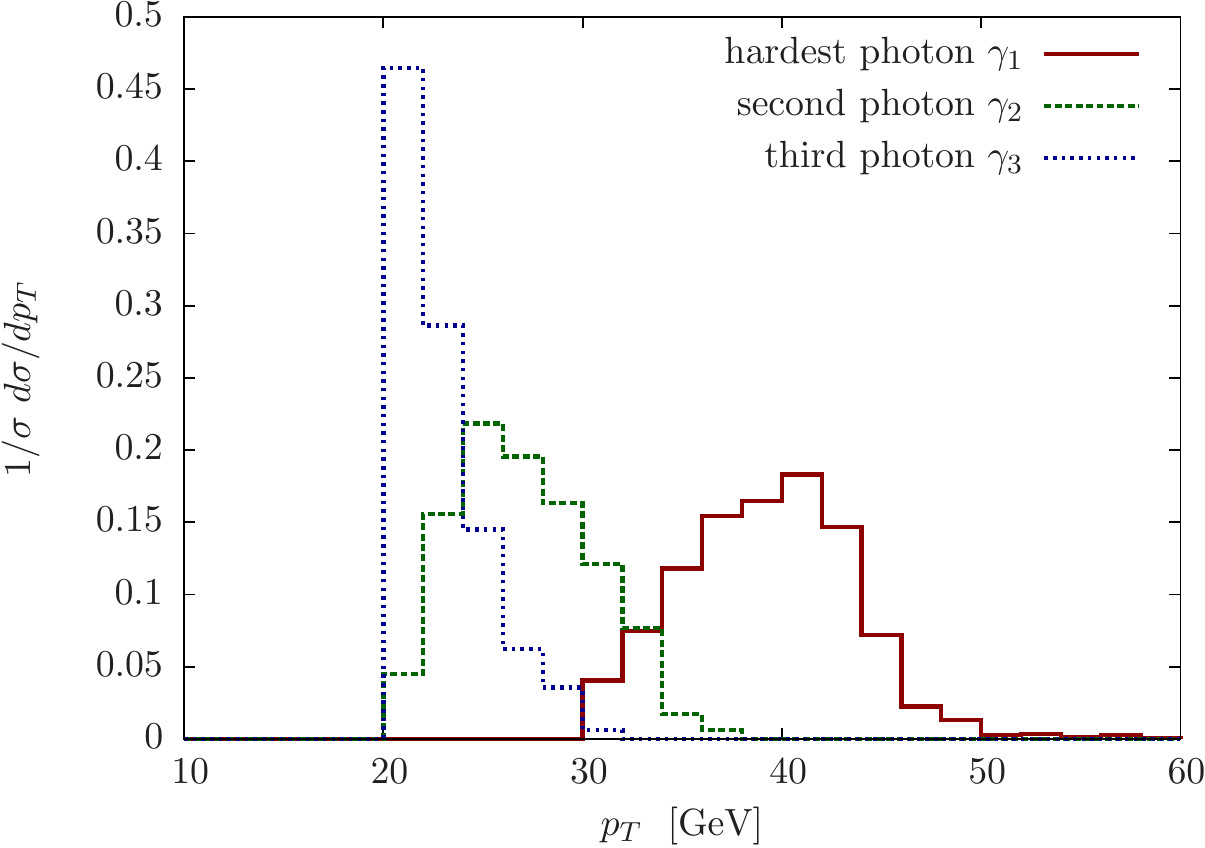}
\includegraphics[width=0.45\textwidth]{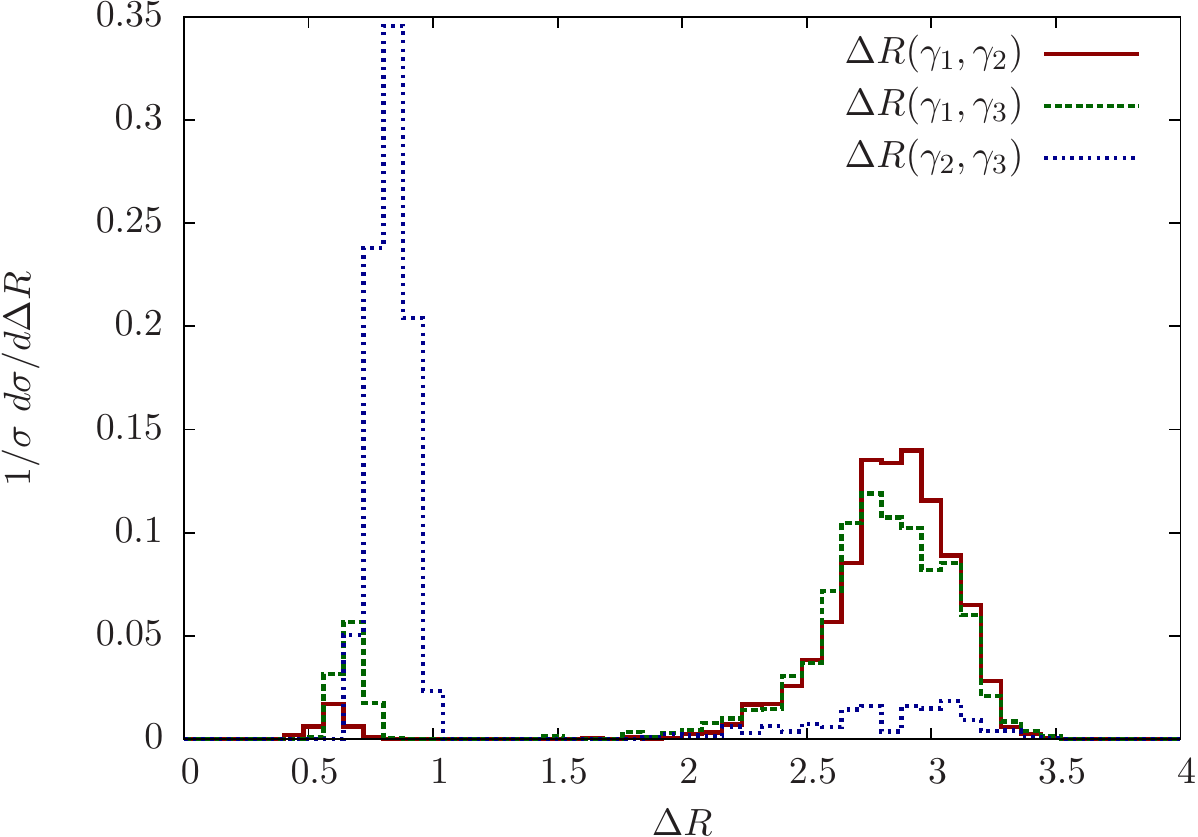} \\
\includegraphics[width=0.45\textwidth]{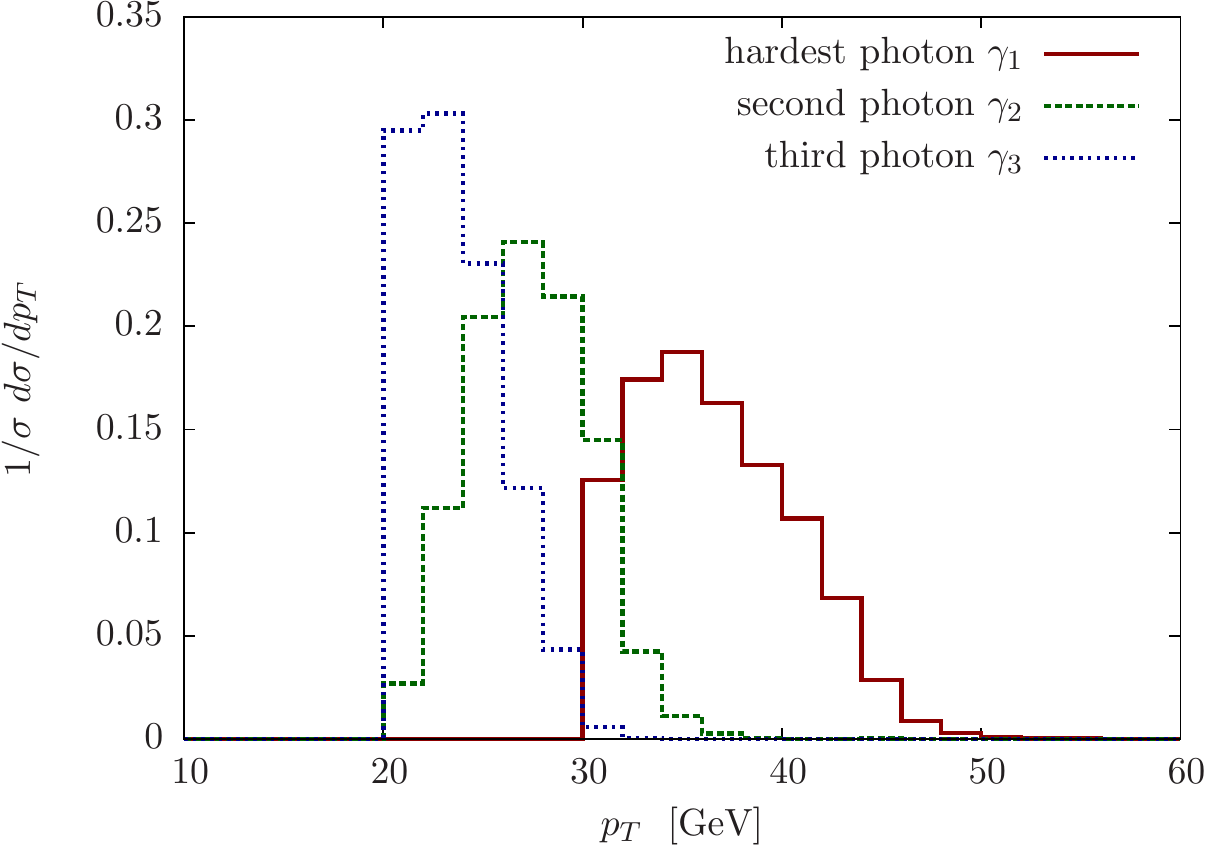}
\includegraphics[width=0.45\textwidth]{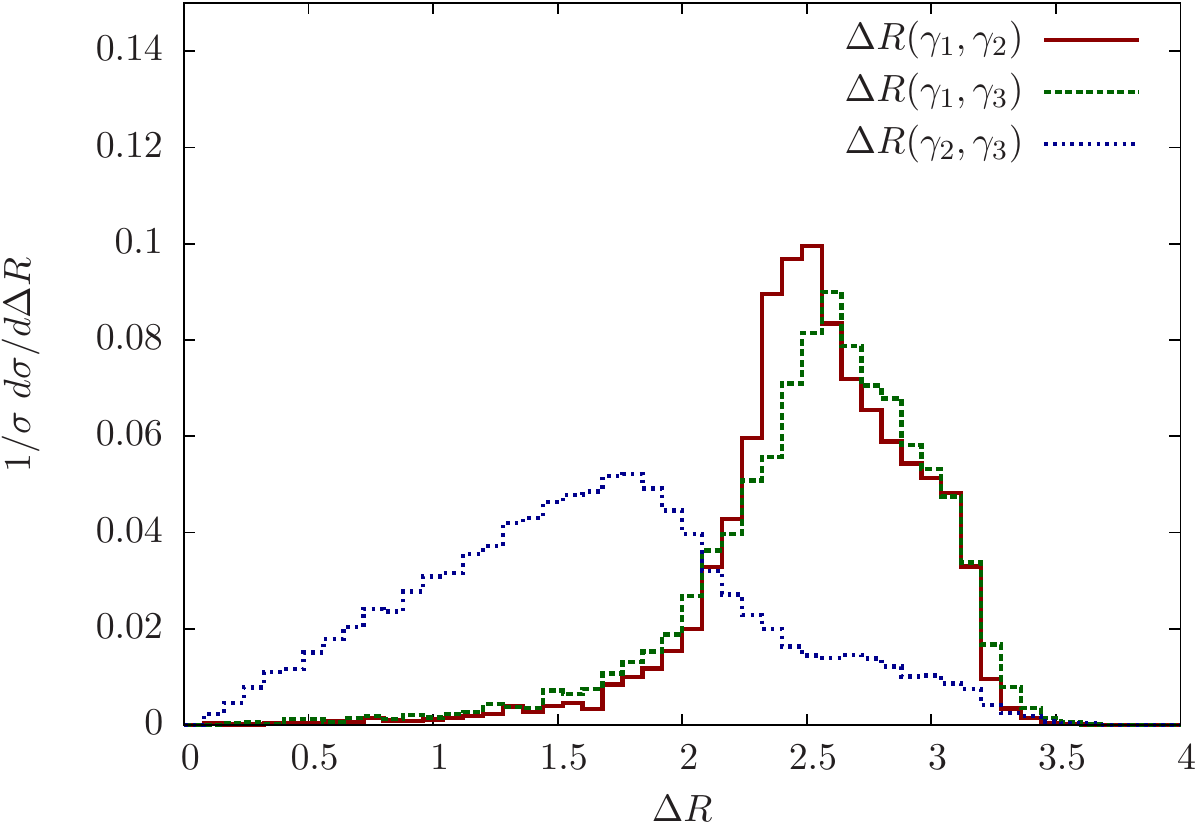} 
\caption{Transverse momentum distributions for the hardest, second and third hardest photons (left). $\Delta R$ separations of the three photons (right). We choose $m_a = 4$ GeV (upper panels), $m_a = 20$ GeV (middle panels) and $m_a=60$ GeV (lower panels).}
\label{fig:dRpT}	
\end{figure}

While the angular separation for two of the photons strongly depends on the mass of $a$, the transverse momentum distribution does not. Hence, to reduce background without biasing our selection towards a specific mass of the axion-like particle, we require for the photons staggered $p_T$ cuts, i.e.
\begin{equation}
\label{eq:pT}
p_{T,\gamma_1} \geq 30~\mathrm{GeV},~~p_{T,\gamma_2} \geq 20~\mathrm{GeV},~~p_{T,\gamma_3} \geq 20~\mathrm{GeV},
\end{equation}
and their invariant mass to be in a window around the $Z$ boson mass,
\begin{equation}
80~\mathrm{GeV} \leq m_{3 \gamma} \leq 100~\mathrm{GeV}.
\end{equation}
Eventually we reconstruct the axion-like particle by requiring that at least one of the three di-photon combinations satisfies
\begin{equation}
\label{eq:ma}
m_a - 3~\mathrm{GeV} \leq m_{\gamma_i \gamma_j} \leq m_a+3~\mathrm{GeV}.
\end{equation}
In Fig.~\ref{fig:dRpT} we show the transverse momentum  distributions and $\Delta R$-separations of the three signal photons for $m_a = 4$, 20 and 60 GeV after the reconstruction steps Eq.~(\ref{eq:pT})-(\ref{eq:ma}). 
For the backgrounds we find generically $\sigma(\gamma jj) \simeq 4 \sigma (\gamma \gamma j) \simeq 9 \sigma(3 \gamma)$, i.e. the much larger inclusive production cross section of $\gamma jj$ is almost, but as a result of the large fake rate for $p_{T,\gamma} \simeq 20$ GeV not quite, reduced to the irreducible $3 \gamma$ background.

We show in Tab.~\ref{tab:reslep} signal and background rates after reconstructing the $Z$ boson and the axion-like resonance. We find that a large range of masses can be excluded with $100~\mathrm{fb}^{-1}$ assuming $\sqrt{s} = 13$ TeV, outperforming existing limits from LEP (see Fig.~\ref{resultleft}). However, while we apply a crude fast-detector simulation to take into account reconstruction efficiencies and fake rates relevant for this process, the signal and background rates have only been calculated at leading-order accuracy. Hence, the statistical significance we quote is plagued by large theory uncertainties. The simple counting experiment we use to evaluate the statistical significance can also be severely affected by normalisation uncertainties and pileup contributions. However, as long as the axion-like particle is a narrow resonance, a side-band analysis, similar to the prominent Standard Model search of a Higgs boson decaying into photons \cite{Aad:2014eha,Khachatryan:2015rxa}, can retain sensitivity irrespective of these problems.
\begin{table}[h]
\begin{center}
\begin{tabular}{c||cc|cc|}
$ m_a$ [GeV] 
& $\sigma_S$ [fb] & $\sigma_B$ [fb] & $S/B$ & $S/\sqrt{B}_{100}$\\
\hline 
$4 $ & 0.0948  & 0.0344 & 2.76 & 5.1 \\
$8$ & 0.0971 & 0.1160 & 0.83 & 2.8 \\
$15$ & 0.0788 & 0.1280 & 0.62 & 2.2 \\
$20$ & 0.0909 & 0.1305 & 0.70 & 2.5 \\
$45$ & 0.1031 & 0.2787 & 0.37 & 2.0 \\
$60$ & 0.0979 & 0.6186 & 0.16 & 1.2 
\end{tabular}
\caption{Signal and background cross sections for $pp \to Z \to a \gamma$ analysis. We assume $g=10^{-4}\,{\rm GeV}^{-1}$ and calculate the final significance for $100~\text{fb}^{-1}$. The cross sections are shown after applying the cuts of Eqs.~(\ref{eq:pT})-(\ref{eq:ma}).}
\label{tab:reslep}
\end{center}
\end{table}

For $m_a < 4$ GeV existing photon reconstruction strategies start to fail since $\Delta R_{\gamma_2 \gamma_3} \lesssim 0.15$. Photons will not be considered isolated anymore and there will be a mistag rate where the two photons induce a shower similar to a single photon. To improve the sensitivity for small masses dedicated reconstruction strategies, e.g. di-/multi-photon taggers \cite{Ellis:2012sd}, would need to be developed.

\section{Conclusions}\label{conclusions}
Relatively light (pseudo-)scalars coupled to two gauge bosons often dubbed axion-like particles (ALPs), are a feature of many extensions of the Standard Model. They are also attractive because of possible connections to dark matter.
$Z$ decays provide a unique opportunity to search for ALPs in the MeV to multi-GeV range. Using data from LEP-I
we have excluded a previously allowed range of masses for ALPs coupled to two photons.
Future precision measurements of $Z$ decays at electron-positron colliders such as FCC-ee promise improvements in the sensitivity by about two orders of magnitude over the current limits. 

We also performed a first analysis on the LHC discovery prospects of light resonances in $3 \gamma$ final states with an invariant mass close to the $Z$ mass, i.e. for unusual and rare $Z$-decays. We find that ATLAS and CMS have a significant discovery potential for ALPs, possibly outperforming LEP.  Hence, a dedicated experimental analysis of this channel would be highly desirable.

\section*{Acknowledgements}
We would like to thank Gustaaf Brooijmans, Babette D\"obrich, Danilo Ferreira de Lima, Philip Harris, Andy Pilkington, Javier Redondo and Marcel Vos for valuable discussions. 
This research was supported in part by the European Commission through the 'HiggsTools' Initial Training Network PITN-GA-2012-316704 and by the Transregio TR33 ``The Dark Universe''.

\end{document}